\documentclass[preprint,12pt]{elsarticle}
\usepackage{graphicx}
\usepackage{amssymb}
\usepackage{amsmath}
\usepackage{lineno}
\usepackage{multirow}
\biboptions{sort&compress}

\journal{arXiv}

\begin{document}

\begin{frontmatter}

\title{Harnessing The Multi-Stability Of Kresling Origami For Reconfigurable Articulation In Soft Robotic Arms}

\author{Joshua Kaufmann and Suyi Li\corref{cor1}}
\fntext[cor1]{Corresponding Author: \texttt{suyil@clemson.edu}}
\address{Department of Mechanical Engineering\\
Clemson University, Clemson, SC}

\begin{abstract}
This study examines a biology-inspired approach of using reconfigurable articulation to reduce the control requirement for soft robotic arms.  We construct a robotic arm by assembling Kresling origami modules that exhibit predictable bistability.  Via switching between their two stable states, these origami modules can behave either like a flexible joint with low bending stiffness or like a stiff link with high stiffness, without requiring any continuous power supply.  In this way, the robotic arm can exhibit pseudo-linkage kinematics with lower control requirements and improved motion accuracy.  A unique advantage of using origami as the robotic arm skeleton is that its bending stiffness ratio between stable states is directly related to the underlying Kresling design.  Therefore, we conduct extensive parametric analyses and experimental validations to identify the optimized Kresling pattern for articulation.  The results indicate that a higher angle ratio, a smaller resting length at contracted stable state, and a large number of polygon sides can offer more significant and robust bending stiffness tuning.  Based on this insight, we construct a proof-of-concept, tendon-driven robotic arm consisting of three modules, and show that it can exhibit the desired reconfigurable articulation behavior.  Moreover, the deformations of this manipulator are consistent with kinematic model predictions, which validate the possibility of using simple controllers for such compliant robotic systems.
\end{abstract}


\end{frontmatter}


\section{Introduction} \label{S:1}

The ongoing advances in bio-mimicry, material science, advanced fabrication, and control theory are enabling us to build genuinely soft robotic arms (or robotic manipulators) that can collaborate with humans in unstructured and dynamic task environments \cite{Trivedi2008, Kim2013, Majidi2014, Laschi2016, Whitesides2018}.  These robotic arms are constructed with soft materials featuring low elastic moduli and high strains before failure, so that they can passively deform their bodies and conform to different objects. This flexibility makes them inherently superior to and safer than the traditional rigid-linked robotic manipulator in human-robot interactions \cite{Sanan2011, Runge2015, Stilli2017}, thus opening up many potential applications in minimal-invasive surgeries \cite{burgner-kahrs_review, Runciman2019, Ranzani2015, Cianchetti2014} and assistive healthcare \cite{Polygerinos2015b, Connolly2016, In2015}.  However, the compliant and continuous nature of these soft robotic arms also imposes significant challenges for effective modeling and control \cite{Camarillo2008a, Trivedi2008a, Renda2014, Thuruthel2018, Webster2010b}.  They usually struggle to achieve a high level of precision regarding their arm configuration and movement control because their soft bodies are high-dimensional and severely under-actuated.  Furthermore, soft materials can exhibit complicated viscoelastic properties with substantial uncertainties.  As a result, inverse kinematics and overall structural shape are difficult to predict, making control tasks such as path planning inaccurate and computationally expensive \cite{Thuruthel2018}.  We are still far away from widely and commercially adopting soft robotics in many aspects of our modern life.

One approach to address the control challenge of soft robotic arms is to decrease their effective degrees of freedom, and nature provides terrific examples of this strategy.  For example, the octopus can generate a quasi-articulated structure with its arm, similar to that of a human, in order to achieve precise point-to-point movement and fetch fast-moving prey within seconds \cite{Sumbre2006}.  The octopus achieves such arm reconfiguration by selectively stiffening sections of its muscles and leaving other sections flexible.  Such reconfigurable structural articulation allows a drastically simplified control by reducing the kinematic degrees of freedom from effectively infinite to a finite amount, thus granting the necessary accuracy to carry out the rapid fetching. To implement this bio-inspired articulation strategy to a soft robotic arm, one must devise a method of localized stiffness tuning to create and activate discrete ``joints'' at different locations.  To this end, researchers have achieved some success by using jamming-based systems \cite{Kim2013a, Cianchetti2013}, low melting point materials \cite{Cheng2010}, and shape-memory polymers \cite{Firouzeh2015a}.  However, these methods are limited due to their complexity and lack of scalability. They also require a continuous energy supply to maintain their respective changes in stiffness \cite{Manti2016}. 

\begin{figure}[]
    \centering\includegraphics[]{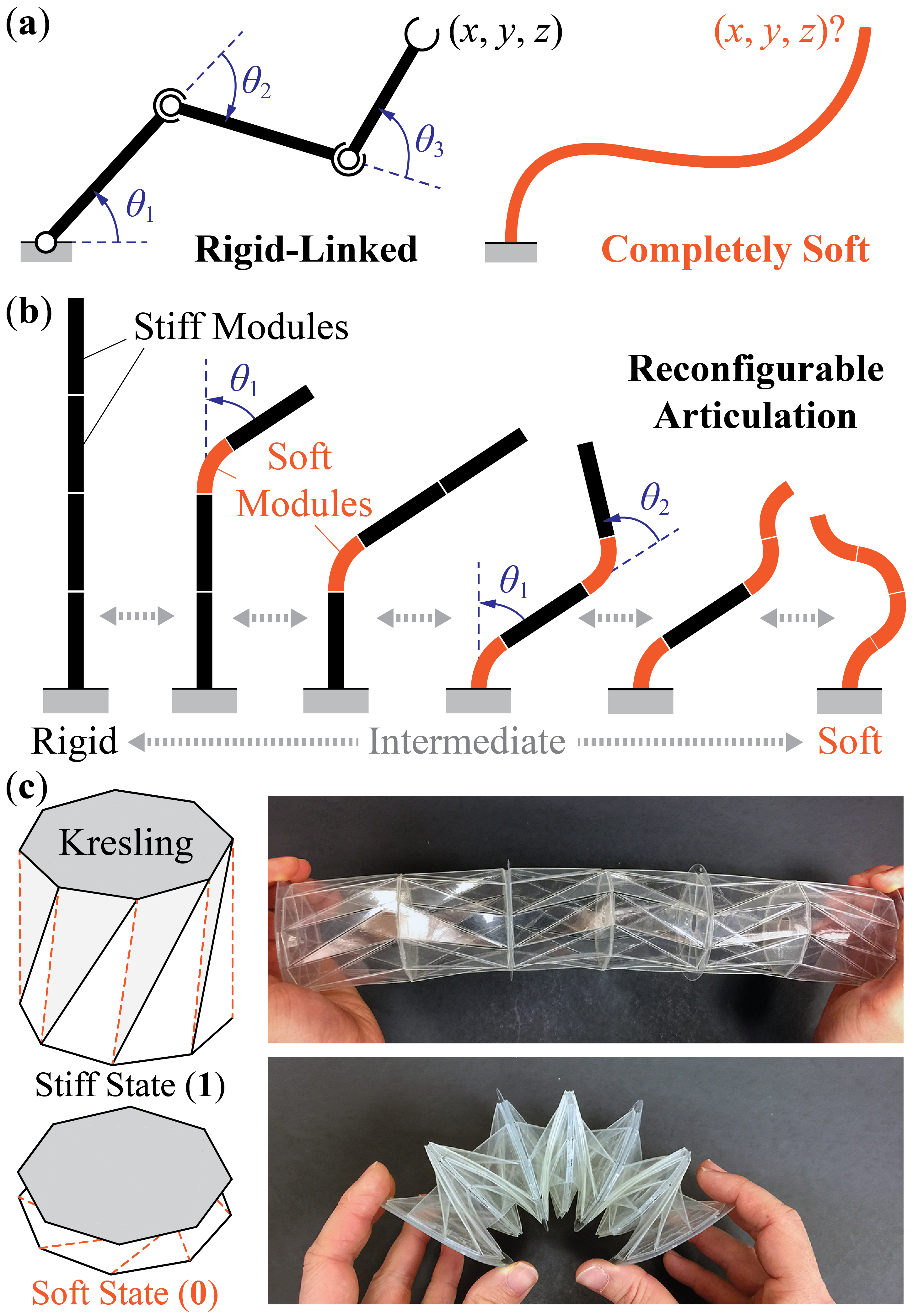}
    \caption{
    An overview of the envisioned reconfigurable articulation in a continuous and compliant robotic arm. a) The current paradigms of the completely rigid or completely soft robotic arm.  b) Different articulations of a reconfigurable arm consisting of four modules. Each module has a relatively stiff stable state and soft state so that this arm can switch from being entirely stiff (left) to entirely flexible (right), as well as to many intermediate configurations with a predictable degree-of-freedom.  Notice that this conceptual robotic arm has 16 $(=2^4)$ unique configurations, and this figure only shows a few representative examples. c) Kresling origami can naturally show the desired switch in bending stiffness between its stable states. In this study, it will be used as the functional skeleton of the robotic arm.
    }
    \label{fig:BigPic}
\end{figure}

In order to achieve the localized stiffness tuning in soft robotic arms in a scalable and energy-passive manner, we seek to analyze and exploit the mechanics of a bistable and cylindrical-shaped origami known as the Kresling (Figure \ref{fig:BigPic}). Kresling origami originates from the buckling and collapsing deformation of a cylindrical shell under compression \cite{Hunt2005}, and it has found many applications in deployable structures \cite{Cai2015, Nayakanti2018, Ishida2017, Li2018} and robotics \cite{Pagano2017, Bhovad2019}.  More importantly, Kresling can fold between two stable equilibria (or stable states) through a coupled longitudinal and rotational motion, and each stable state possesses unique mechanical properties according to its folding geometry.  Such bistability enables a method of binary bending stiffness tuning. Therefore, we can construct a soft robotic arm by serially connecting Kresling cells (or modules) and create joint(s) at any desired locations bDy switching these cells between their stable states (Figure \ref{fig:BigPic}).  This approach is unique in that the localized stiffness tuning is embodied in the skeleton of the robot arm itself, and the mechanics are scalable because they are derived primarily from folding geometry.

The objective of this study is twofold.  First, we examine the correlation between Kresling origami design and the bending stiffness ratio between its two stable states. A significant and robust change in bending stiffness is the key to successful articulation in the soft robotic arm.  To this end, we employ a nonlinear bar-hinge model, together with experimental validation, to identify the optimized Kresling pattern design.  Parametric analysis findings indicate that a higher angle ratio, a smaller resting length at the contracted stable state, and a large number of polygon sides result in a considerable change in bending stiffness.  In particular, the reorientation of the triangular facets between stable states plays a crucial role.

Based on these insights, the second objective of this study is to validate the feasibility of manipulator articulation via  multi-stability.  To this end, we construct a proof-of-concept, tendon-driven manipulator consisting of three Kresling modules, and show that it can exhibit the desired reconfigurable articulation behavior.  Moreover, the deformation of this manipulator is consistent with the kinematic model prediction, which validates the feasibility of using simple controllers for such compliant robotic systems.

In what follows, section 2 of this paper briefly reviews the design of the generalized Kresling origami, which is a variation of the classical Kresling pattern to accommodate the kinematic requirement of robotic manipulation.  Section 3 details the parametric analysis and experimental testing of the bending stiffness of a Kresling origami module.  Section 4 details the results of the robotic arm prototype.  Finally, section 5 concludes this paper with a summary and discussion.  The results of this study will lay down the foundation for constructing a new family of hybrid soft robotics arms that are both flexible in operation and precise in motion capability.

\section{Design and Construction of Generalized Kresling} \label{S:2}

\begin{figure}[]
    \centering\includegraphics[]{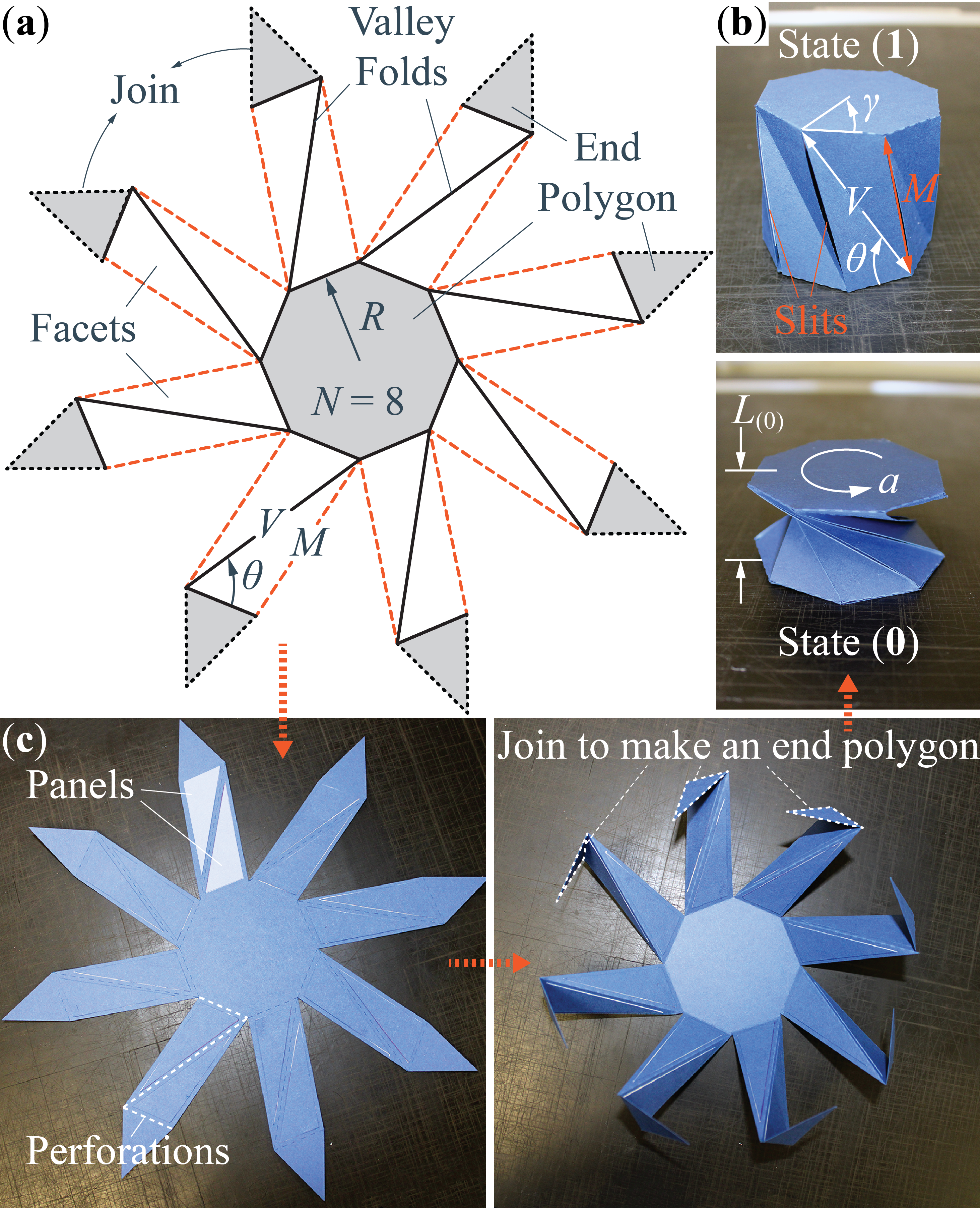}
    \caption{
    Design and construction of a paper based Kresling origami cell. In this design $N=8$, $\lambda=0.8$, $L_{(0)}=30$ mm, and $R=30$ mm (a) The origami crease pattern showing different design parameters.  Notice that the triangular tips should be joined together to form an end polygon. (b) The completed Kresling cell at two different stable states.  Notice that the mountain folds on the side are open slits by design.  (c) The intermediate states of folding Kresling.  Triangular panels are attached to reinforce the facets to improve overall bistability.
    }
    \label{fig:Design}
\end{figure}

A generalized Kresling origami cell consists of a group of triangular facets connected by two polygon end surfaces (Figure \ref{fig:Design}).  Once assembled, the Kresling cell takes a twisted polygonal prism shape.  The convex creases (or mountain creases) on its side are open slits by design in order to ensure flexibility and robustness during folding and bending \cite{Bhovad2019}.  A Kresling cell can settle into an extended stable state (referred to as ``state (1)'' for simplicity hereafter) or a contracted stable state (aka. ``state (0)'').  The bistability of Kresling origami originates from its non-rigid-foldable nature. That is, the triangular facets are flat and undeformed at the two stable states, but must deform during the folding transition between these two states.  If these triangular facets were strictly rigid, the Kresling segment would be unable to fold. 

Four independent design parameters can fully define the crease pattern of the generalized Kresling cell.  They are 1) the number of polygon sides ($N$), 2) radius ($R$), 3) resting length at the contracted stable state ($L_{(0)}$), and 4) an angle ratio ($\lambda$).  Here, $L_{(0)}$ is the variable that differentiates the generalized Kresling origami from conventional Kresling.   The traditional Kresling has a zero length by definition at the contracted state (a property often known as ``flat-foldable''). However, a zero resting length would prevent any kinematic freedom for bending---an essential requirement for robotic arm applications.   Therefore, we generalized the Kresling design with a non-zero resting length at the state (0) to provide the freedom for bending so that the Kresling cell can work like a revolute ``joint'' in the pseudo-articulated structure.  The angle ratio ($\lambda$) influences the strength of the Kresling bistability.  The Kresling becomes bistable when $0.5<\lambda<1$.  Moreover, the higher the angle ratio, the stronger the bistability becomes in that one needs to apply a higher force to fold the Kresling between two stable states \cite{Bhovad2019}. 

Once we prescribe the aforementioned design variables, the triangular facets can be defined as

\begin{equation}
    V =\sqrt{4R^2\cos^2\left(\gamma-\lambda \gamma \right) + L_{(0)}^2},
\end{equation}

\begin{equation}
    M =\sqrt{P^2+V^2-4PR\cos\left(\gamma-\lambda \gamma \right)\cos\left(\lambda \gamma \right)},
\end{equation}

\begin{equation}
    \theta =\cos^{-1}\left(\frac{P^2+V^2-M^2}{2PV} \right),
\end{equation}
where $\gamma$ $(=\pi/2-\varphi)$ is the angle between the diagonal and side of the end polygon, $P$ $(=2R\sin\varphi)$ is the end polygon side length, and $\varphi=\pi/N$.

To calculate the resting length of the generalized Kresling origami at its extended stable state (aka. $L_{(1)}$), we first introduce $\alpha$---the relative rotation angle between the top and bottom end polygon---as the independent variable that describes the folding motion. Moreover, we assume the end polygons are rigid, and the valley creases do not change their length.  In this way, facet deformation in the Kresling cell during folding can be approximated by the shortening of mountain creases, and we can calculate the current mountain crease length ($m$) as well as the overall Kresling cell length ($l$) as functions of $\alpha$ \cite{Bhovad2018, Bhovad2019}:

\begin{equation}
    l(\alpha) = \sqrt{L_{(0)}^2+2R^2\left[\cos(\alpha+2\varphi)-\cos(\alpha_{(0)}+2\varphi)\right]},
\end{equation}

\begin{equation}
    m(\alpha) = \sqrt{2R^2(1-\cos(\alpha))+l^2}.
\end{equation}

Based on the equations above, one can find the extended stable state (1) by solving $m(\alpha)=M$ and fully determine the external geometry of the generalized Kresling origami cell at its two different stable states.  

\section{Bending Stiffness tuning of Kresling Modules} \label{S:3}

We use both numerical modeling and experimental testing to examine the bending stiffness tuning of Kresling modules between their two stable states, as well as the correlations between this stiffness tuning and underlying origami design.

\subsection{Methods}
\begin{figure}[]
    \hspace{-0.5in}
    \includegraphics[]{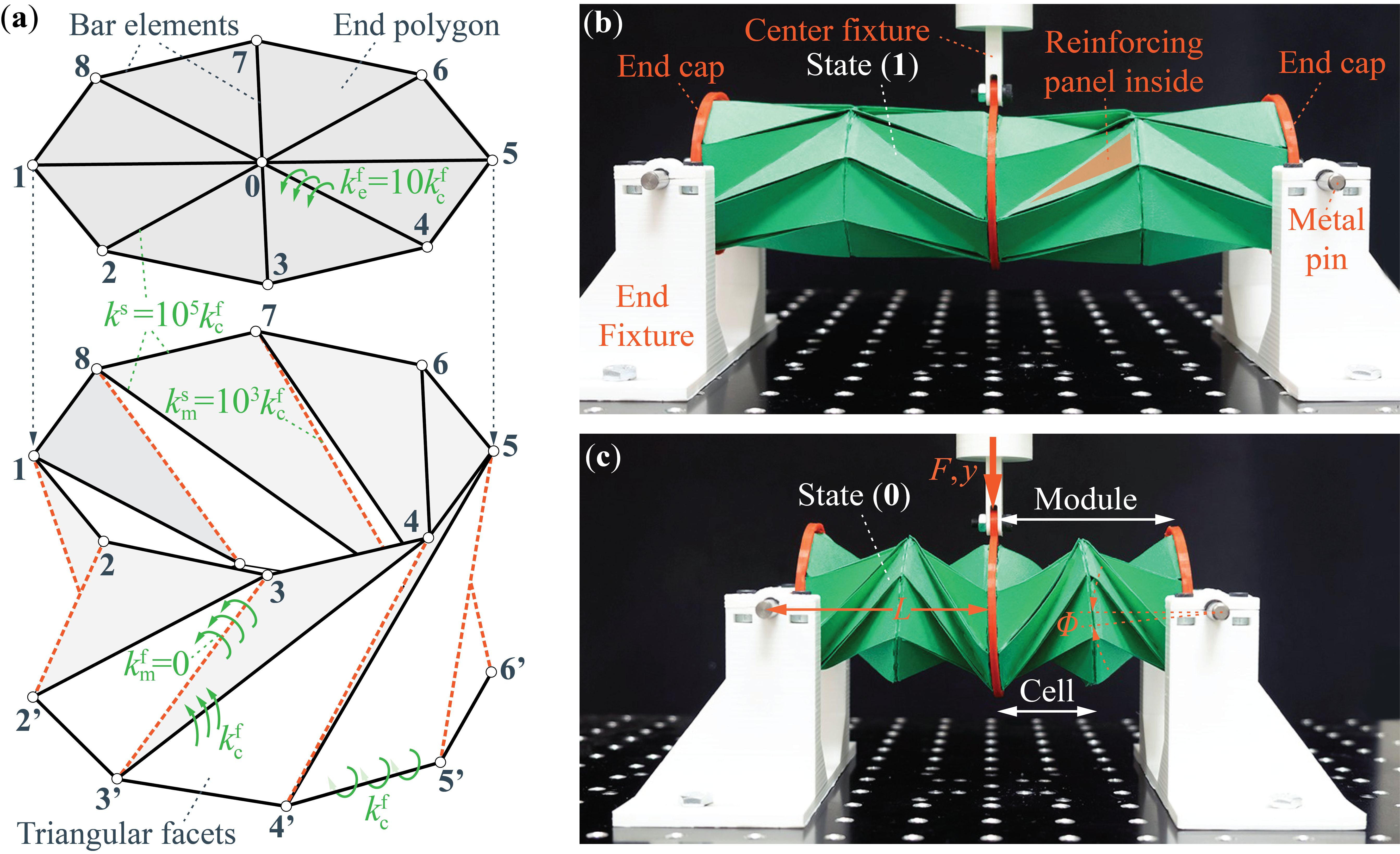}
    \caption{
    The numerical and experimental methods used for testing the bending stiffness of Kresling origami. (a) An illustration of the nonlinear bar-hinge methods, where solid lines represent the stretchable bar elements, and small circles represent the pin-joints. The different bar rigidity and folding/bending stiffness per unit length are highlighted for clarity. (b, c) The three-point bending tests at two different stable states. Notice the definition of a Kresling ``module'' in (c), which is an assembly of two Kresling cells of the same design but different chirality.
    }
    \label{fig:BendTest}
\end{figure}

For numerical modeling, we adopt a nonlinear bar-hinge approach that transforms the Kresling origami into a pin-jointed bar-hinge (or truss-frame) structure.  This approach uses stretchable bar elements to represent the creases and adds additional torsional spring coefficients to approximate the crease folding and facet bending stiffnesses (Figure \ref{fig:BendTest}(a)).  These simplifications result in a reduced-order model capable of analyzing the folding kinematics and principle deformations of Kresling origami without incurring the expensive computational cost associated with three-dimensional finite element simulations \cite{Li2018}.   Interested readers can refer to supplemental materials (Appendix 1) for the fundamentals of this bar-hinge approach, and relevant literature for further details \cite{Liu2017c, Liu2019}. 

The overall stiffness of the equivalent bar-hinge structure has two components. One comes from the stretching of the bar elements and the other from the folding (or bending) between adjacent triangular facets defined by these bar elements.  Therefore, it is crucial to assign appropriate elastic properties to this bar-hinge system so that it can accurately represent the mechanics of the Kresling module.  In this study, we assume the axial bar rigidity ($k^\text{s}$) and folding/bending torsional stiffness \emph{per unit length} ($k^\text{f}$) are all constant so that the Kresling nonlinearity originates from the large amplitude deformation during folding only.

There are two different types of torsional spring coefficients in this Kresling bar-hinge system.  The first type is the torsional stiffness per unit length of the origami creases ($k^\text{f}_\text{c}$), which applies to the valley creases on the side (e.g., 2'-3 and 3'-4 in Figure \ref{fig:BendTest}(a)) and the creases between end polygons and triangular facets (e.g., 2-3 and 2'-3').  The magnitude of this stiffness is experimentally measured to be 0.047 N/radian on average (see Appendix 2 in supplemental materials for test details).  The second type of torsional stiffness per unit length ($k^\text{f}_\text{e}$) applies to the end polygon (e.g., along 0-1 and 0-2).  Its magnitude is assumed to be an order of magnitude higher than the crease torsional stiffness because they represent the polygon material bending ($k^\text{f}_\text{e}=10k^\text{f}_\text{c}$).  It is worth noting that the torsional spring coefficient of the mountain creases (e.g., 2-2' and 3-3') is zero due to the open slit design.

Besides the torsional spring coefficients, we assume the same axial rigidity for all bar elements in that $k^\text{s}=10^5k^\text{f}_\text{c}$ \cite{Liu2017c}.  However, one exception is the bar elements along the mountain creases on the Kresling side (e.g., 2-2' and 3-3').  The axial rigidity of these bar elements is two orders of magnitude lower due to the slit cut design (aka. $k^\text{s}_\text{m}=10^3k^\text{f}_\text{c}$).

We also fabricated paper-based prototypes of the Kresling module and conducted a three-point bending test (Figure \ref{fig:Design}(b, c)). To fabricate a Kresling cell, we first create the Kresling geometry in a CAD program and convert it into a vectorized image file.  This file is then sent to a cutting plotter (Graphtec FCX4000-50ES) that can accurately perforate the crease lines and cut the Kresling cells out of a large piece of thick paper (Daler-Rowney Canford 150 gsm). We then manually fold these cells and assemble them into Kresling modules for testing.   It is worth noting that the triangular facets have reinforcement panels attached inside to increase their bending stiffness, thus increasing the bistability strength \cite{Bhovad2019}.

A single Kresling cell, however, shows twisting in addition to longitudinal deformation when it folds from one stable state to the other, and this twisting is undesirable for the robotic manipulation purpose.  Therefore, we construct a Kresling ``module'' by combining two kresling cells of the same design but opposite chirality (Figure \ref{fig:BendTest}(c)).  In this way, the two end polygons of a module do not rotate with respect to each other.  We secure an assembly of two identical Kresling modules to the universal testing machine (ADMET eXpert 5601) and fix them at either their extended (1) or contracted (0) stable state (Figure \ref{fig:BendTest}(b, c)). We then apply a 5 mm downward displacement, with a rate of 0.1 mm/sec, at the center (aka. a three-point bending test).  In this way, the effective bending stiffness of a Kresling module is 

\begin{equation} \label{eq:KB}
K_B=\frac{M}{\phi}=\frac{FL}{\tan^{-1}\left(y/L\right)},
\end{equation}
where $M$ is the applied moment, $\phi$ is the rotation angle, $F$ is the reaction force, $y$ is the downward displacement, and $L$ is the distance between applied force and rotation axis of the Kresling module.  To characterize the performance of Kresling module, we define a ``bending stiffness ratio'' as the ratio of bending stiffness at the stiff stable state (1) to that at the soft stable state (0).

\begin{figure}[]
    \hspace{-0.5in}
    \includegraphics[]{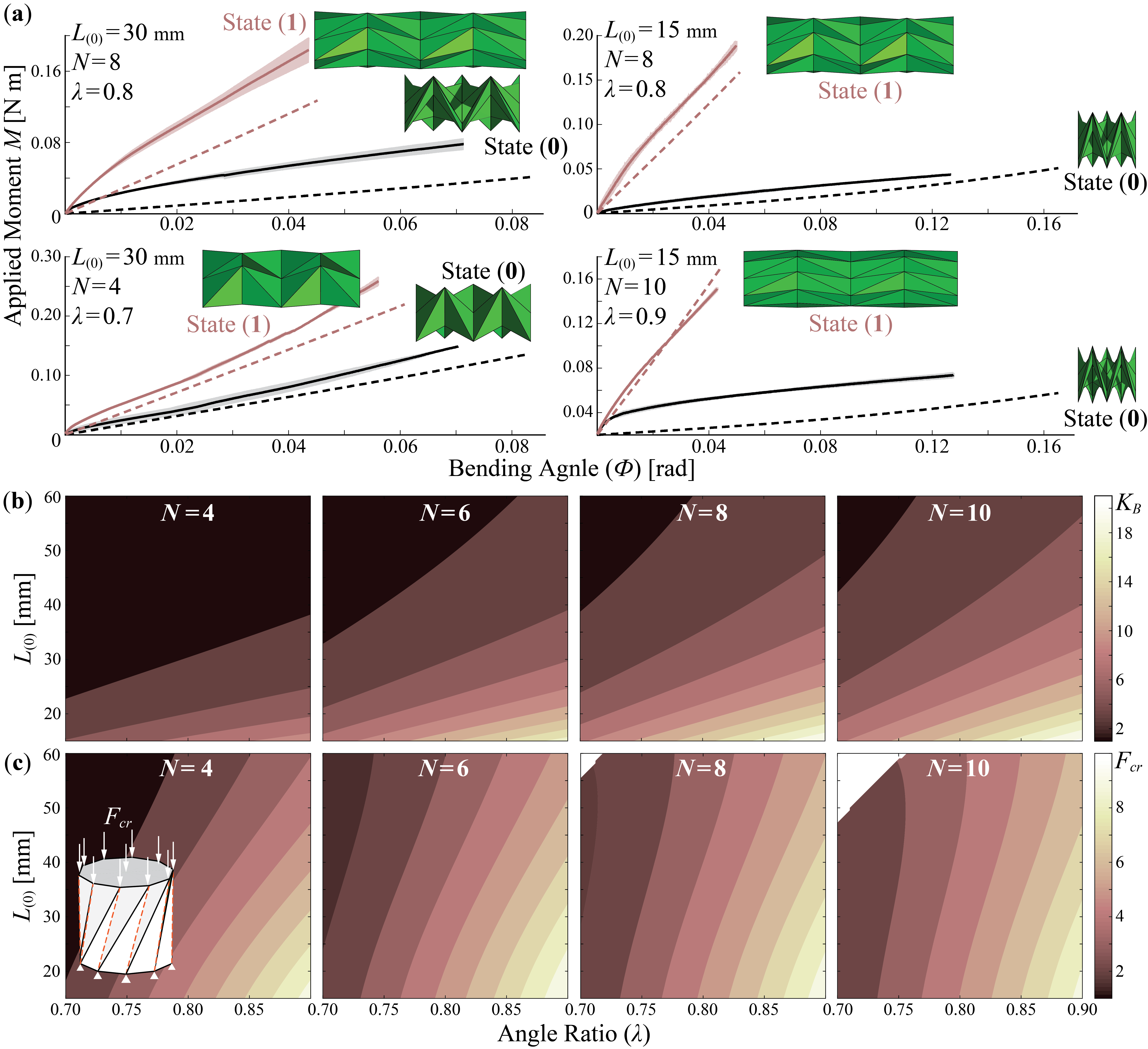}
    \caption{
    Bending stiffness change between the two stable states of Kresling modules with different designs.  (a) Experiment results of the three-point bending tests (solid lines) and the corresponding numerical predictions (dashed lines).  The shaded bands are the standard deviation of three loading cycles. (b, c) Parametric analysis results of the bending stiffness ratio and axial snap-through force (in N), respectively.
    }
    \label{fig:BendResult}
\end{figure}

\subsection{Results}
Figure \ref{fig:BendResult}(a) summarizes the external moment-bending angle relationships of four Kresling module assemblies based on different crease designs.  The results indicate that these origami modules can show a significant change in bending stiffness as they switch from one stable state to the other.  Moreover, this change in bending stiffness is directly related to the underlying origami design.  Overall, the bar-hinge model predictions correlate with the experimental results well.  However, some discrepancies exist at the beginning stage of these tests.  Compared to the model predictions, the experiment results show a stronger nonlinearity when the external load is small.  This discrepancy probably originates from the fact that the Kresling module prototypes have to be compressed slightly before testing so that their initial length equals to the theoretically predicted resting length at the two stable states (aka. $L \approx L_{(0)}$ or $L_{(1)}$).  This small compression generates some initial stress in the structure.  Regardless, the Kresling test samples show a close-to-linear behavior as the bending angle $\theta$ increases, and the analytical and experimental results agree well with each other in terms of the slope of these moment-angle curves (Table \ref{tab:KB}).

\begin{table}
    \renewcommand{\arraystretch}{1.25}
    \centering
    \begin{tabular}{c|c|c|c|c} 
    \hline
    Design & Test & $K_B$ at (1) & $K_B$ at (0) & $K_B$ Ratio\\
    \hline
    \multirow{2}{8em}{$N=8$, $\lambda=0.8$, $L_{(0)}=30$ mm} 
        & Experiment    & 3.50 $\pm$ 0.29 & 0.80 $\pm$ 0.07 & 4.9 \\ 
        & Bar-Hinge     & 2.79            & 0.48            & 5.8 \\ 
    \hline
    \multirow{2}{8em}{$N=8$, $\lambda=0.8$, $L_{(0)}=15$ mm} 
        & Experiment    & 3.59 $\pm$ 0.17 & 0.29 $\pm$ 0.01 & 12.3 \\ 
        & Bar-Hinge     & 3.08            & 0.27            & 11.4 \\ 
    \hline
    \multirow{2}{8em}{$N=4$, $\lambda=0.7$, $L_{(0)}=30$ mm} 
        & Experiment    & 4.31 $\pm$ 0.09 & 2.20 $\pm$ 0.19 & 2.0 \\ 
        & Bar-Hinge     & 3.59            & 1.61            & 2.2 \\ 
    \hline
        \multirow{2}{8em}{$N=10$, $\lambda=0.9$, $L_{(0)}=15$ mm} 
        & Experiment    & 2.88 $\pm$ 0.03 & 0.25 $\pm$ 0.02 & 11.5 \\ 
        & Bar-Hinge     & 3.35            & 0.20            & 16.8 \\ 
    \hline
    \end{tabular}
    
    \caption{Summary of the three-point bending test results on four different Kresling module samples. The unit of bending stiffness $K_B$ is [N m/rad].}
    \label{tab:KB}
    
\end{table}

Based on the experimentally validated model, we conduct a parametric analysis to obtain a comprehensive understanding of how we can tune the bending stiffness by tailoring the underlying Kresling design (Figure \ref{fig:BendResult}(b)).  In this analysis, we keep the Kresling module radius $R$ as a constant at 30 mm because it is usually determined by specific application requirements.  We vary the magnitude of the other three independent design parameters such that the polygon side number $N$ is 4, 6, 8, or 10, the angle ratio $\lambda$ is between 0.7 and 0.9, and the resting length at the contracted stable state $L_\text{(0)}$ is between 15 mm and 60 mm.  These parameter ranges are chosen carefully according to the fabrication constraints.  The numerical simulation results conclude that, regardless of the number of polygon sides, maximizing the angle ratio $\lambda$ or minimizing the contracted length $L_\text{(0)}$ can lead to a higher bending stiffness ratio.  In other words, if the Kresling design possesses stronger bistability and is closer to the traditional design (aka. with zero $L_\text{(0)}$), it will have a higher bending stiffness ratio.  Increasing the base polygon sides ($N$) can also increase the bending stiffness ratio.  However, this benefit becomes marginal when $N$ becomes bigger than 8. 

\begin{figure}[]
    \centering \includegraphics[]{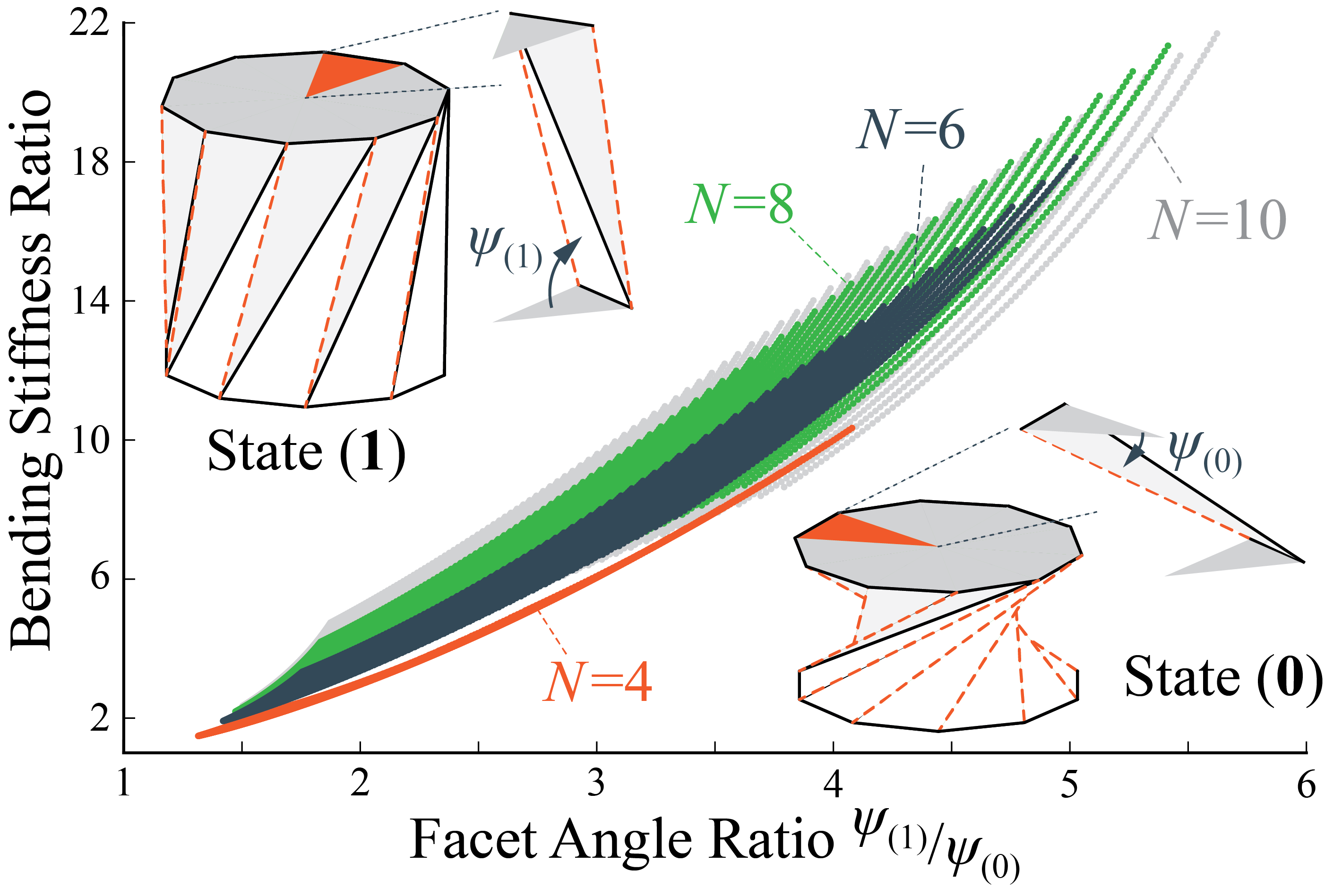}
    \caption{
    The correlation between the facet angle ratio and bending stiffness ratio.  Here, each point represent a unique Kresling origami design used in the parametric analyses in Figure \ref{fig:BendResult}.
    }
    \label{fig:PanelAngle}
\end{figure}

Careful inspection of the Kresling origami geometry can reveal the physical principles that underpin this stiffness change.  At the extended stable state (1), the triangular facets in the Kresling origami align close to parallel to the longitudinal axis of the cylindrical-shaped module, so the overall bending stiffness is relatively high and dominated by the facet stretching.  However, at the contracted state (0), the triangular facets are orientated close to perpendicular to the longitudinal axis, so the bending stiffness is low and dictated by the crease folding.  To validate this causality, we define a ``facet angle ratio'' as the ratio of dihedral angels between the triangular facet and base polygon between the two stable states.  Figure \ref{fig:PanelAngle} indicates that a higher facet angle ratio directly creates a higher bending stiffness ratio.

Finally, it is also essential that the stiffness tuning of the Kresling module is robust against external disturbances.  That is, once the Kresling module settles into the targeted stable state, it should remain in this state so that external forces (e.g., from the payloads) will not create any unintentional switch.  To evaluate this robustness, we use the bar-hinge model to calculate the axial force required to switch the Kresling model from the extended state to the contracted state (Figure \ref{fig:BendResult}(c); it is worth noting that pure bending would not create any switch between stable states).  Interestingly, the parametric analysis shows that the Kresling designs with a higher bending stiffness ratio also requires a large axial force to be compressed from state (1) to (0), thus showing a more robust bistability.  Based on the results discussed above, we choose $N$ = 10, $\lambda$ = 0.9, and $L_\text{(0)}$=15 mm as the optimal Kresling module design to construct a proof-of-concept robotic arm with reconfigurable articulation (as we detail in the following section).

\section{Robotic Arm Articulation}

Using the optimal Kresling pattern design, we construct a proof-of-concept robotic arm using three identical modules (or 6 Kresling cells of alternating chirality).   However, instead of using thick paper like in the bending stiffness study, we use a more robust, layered polymer sheet construction for the robotic arm. It is worth noting that since the correlation between Kresling design and bending stiffness tuning is dictated by the folding geometry, insights from the previous section apply regardless of the materials selected.

\begin{figure}[]
    \centering \includegraphics[]{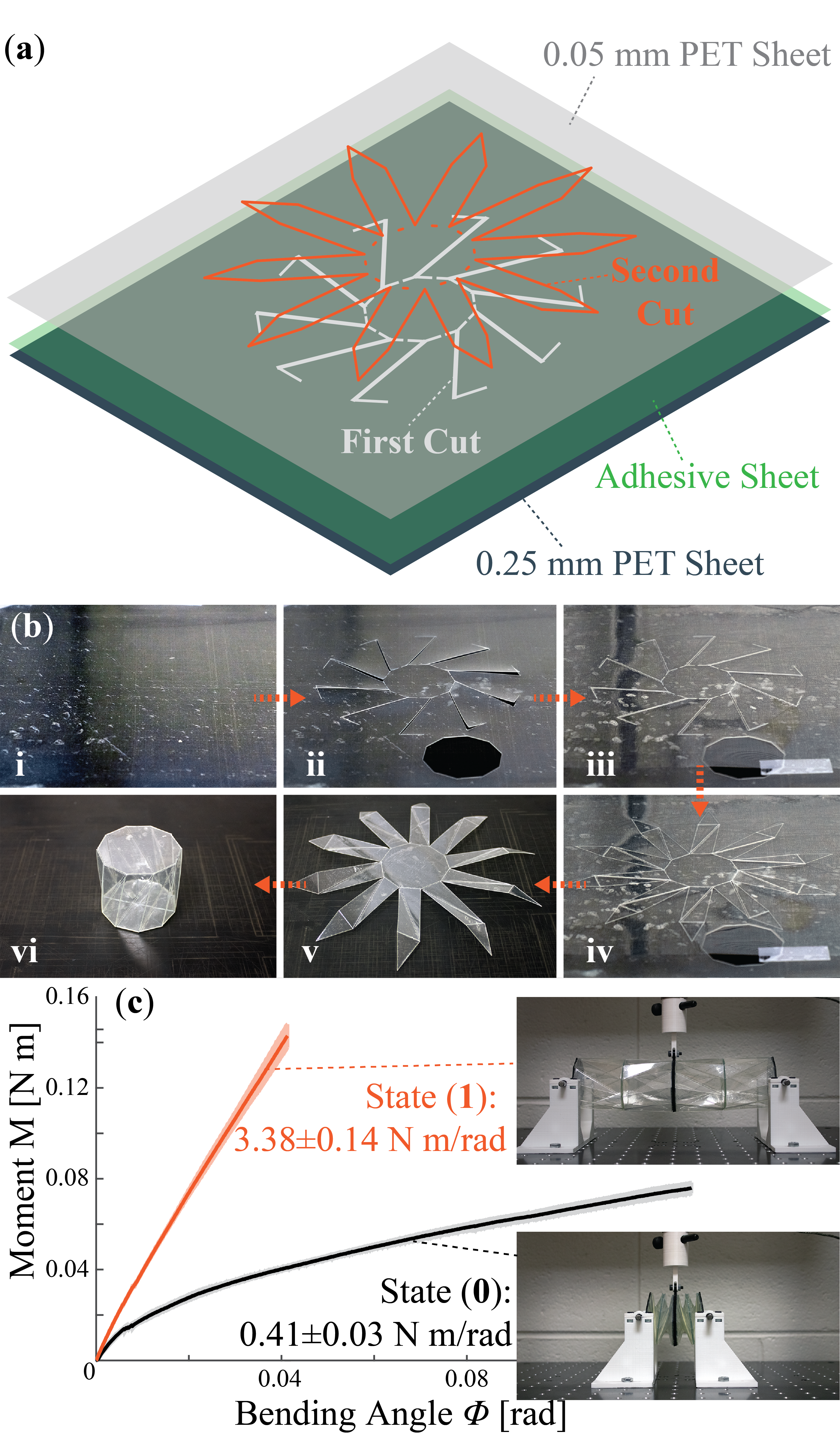}
    \caption{
    Fabrication of the PET plastic-based Kresling cell for constructing the proof-of-concept, robotic manipulator.  (a) The layered construction method involves two plastic sheets, one adhesive sheet for bonding, and two different cuts.  (b) The fabrication sequence, from top-left and clockwise, i: bond the 0.25 mm PET sheet and adhesive sheet and secure them on the cutting plotter;  ii: perform the first cut, notice that an end polygon piece is also cut out; iii: attach the 0.05 mm PET sheet; iv: perform the second cut; v: remove the cut Kresling pattern from the cutting plotter; vi: manually fold the Kresling and attach it to the end polygon piece. (c) the three-point bending test results.
    }
    \label{fig:RobotFab}
\end{figure}

In the plastic-based Kresling, a layer of 0.25 mm thick polyethylene terephthalate (PET) sheet serves as the base material for the facets, and an additional layer of 0.05 mm thick PET sheet is used for the flexible creases.  F9460PC adhesive transfer tape bonds these polymer sheets seamlessly (see Figure \ref{fig:RobotFab} for fabrication details).   A three-point bending test shows that the plastic Kresling module can exhibit a bending stiffness ratio of 8.24 between its two stable states.  

\begin{figure}[]
    \hspace{-0.6in}
    \includegraphics[]{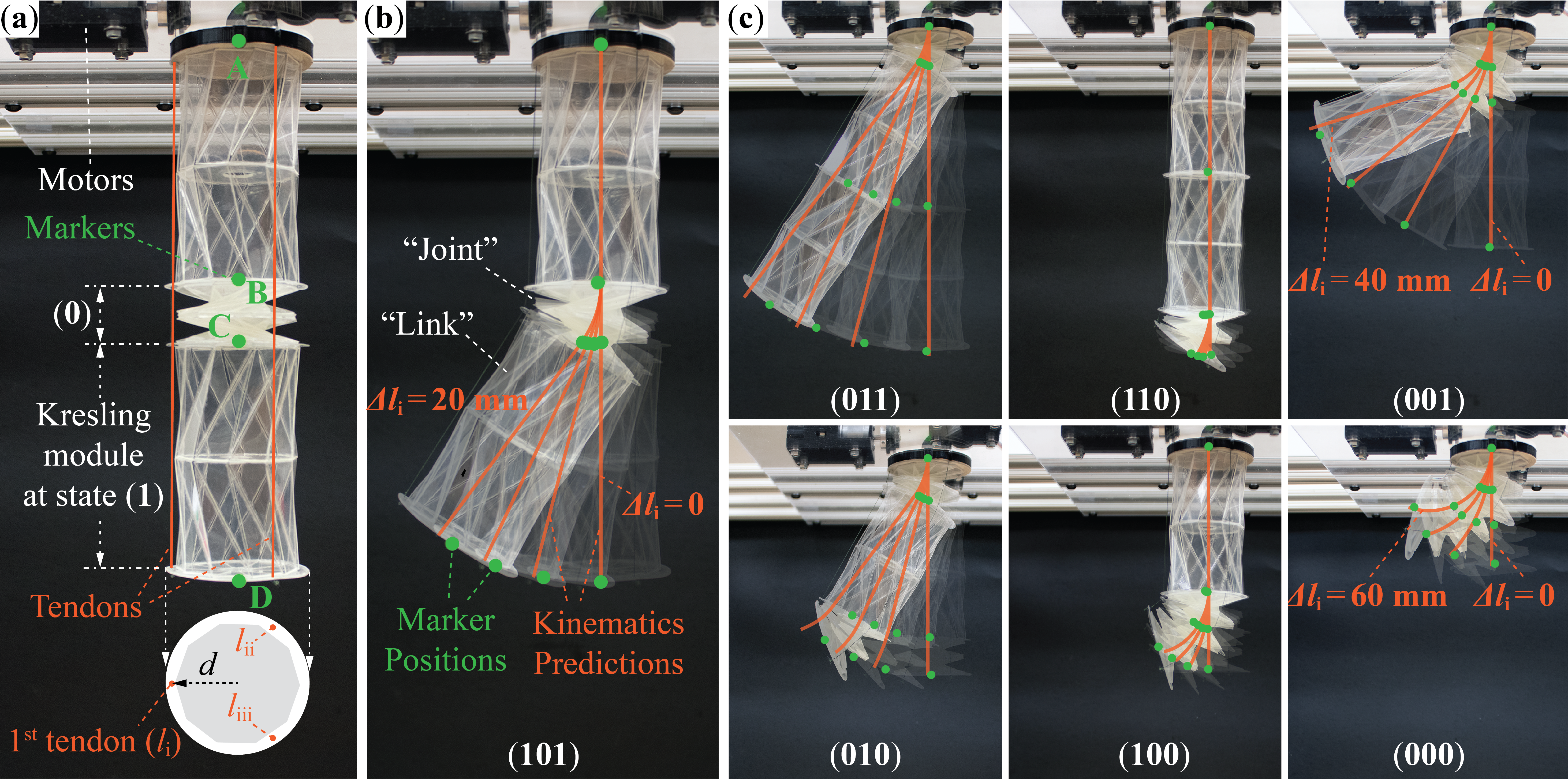}
    \caption{
    Proof-of-concept test of the Kresling robotic arm with reconfigurable articulation. a) Test setup that shows the arm with three identical Kresling modules. The three motor-driven, evenly-spaced tendons i, ii, and iii are highlighted.  We also attached green-colored markers A, B, C, and D at the endpoints of each Kresling module to facilitate measurement. b) A composite image that shows the robotic arm deformation---with the (101) configuration---at different tendon actuation levels.  The predicted arm deformations based on the kinematic model are plotted directly on-top-of the composite image as solid lines, which show good agreement with the green marker positions.  c) Similar composite images showing the robotic arm deformations and corresponding kinematic predictions at all other stable configurations. In these composite images, the tendon actuation $\Delta l_\text{i} =$ $0$, $8n_\text{(0)}$, $14n_\text{(0)}$, and $20n_\text{(0)}$mm, respectively, where $n_\text{(0)}$ is the number of Kresling modules at state (0) corresponding to the current articulation configuration.  For example, $n_\text{(0)}=1$ for (101), $n_\text{(0)}=2$ for (001), and $n_\text{(0)}=3$ for (000).
    }
    \label{fig:RobotExp}
\end{figure}

We use three evenly spaced, motor-driven tendons routed through the Kresling robotic arm skeleton as the driving mechanism. Figure \ref{fig:RobotExp} summarizes the arm deformation at different configurations.  With three Kresling modules, the robotic arm can settle into 8 ($=2^3$) different configurations.  For clarity, we label these configurations by $(ijk)$ where $i$, $j$, and $k$ can take the value of either 0 or 1, representing the current stable state of three Kresling modules from the robotic arm base to the tip, respectively.   Obviously, the $(111)$ configuration is immobile since all Kresling modules are in the stiffer stable state $(1)$, and the $(000)$ configuration is entirely soft, just like the more conventional soft manipulators.  In this proof-of-concept test, we manually set the Kresling robotic arm into the desired articulation configuration, and the methods of automatic reconfiguration will be a subject of a follow-up study. 

The robotic arm deformations for the other six configurations validate the feasibility of reconfigurable articulation.  That is, Kresling modules settled at the stable state $(0)$ act as joints, while those at state $(1)$ act as links (Figure \ref{fig:RobotExp}(b,c)). To further illustrate the feasibility of using simple controllers for the articulated Kresling robotic arm, we attach green-colored markers at the end points of each Kresling module, take high resolution images of the robotic arm at different tendon actuation inputs, use image processing software to accurately measure the arm deformation by tracking the position of markers, and finally compare these measurements to analytical predictions based on a kinematic model. In Figure \ref{fig:RobotExp}(b,c), the kinematic predictions are plotted directly on top of the corresponding robotic arm images for comparison.  

The kinematic model in this study adopts the Denavit-Hartenberg (DH) convention, which represents the transformation of coordinates from the reference frame attached to one Kresling module to another \cite{Spong2006}.  By defining four different reference frames at the four endpoints of Kresling modules (aka. $A$, $B$, $C$, and $D$ in Figure \ref{fig:RobotExp}(b)), we can describe the transformation from a reference frame $A$ to frame $B$ as:

\begin{equation}
\textbf{H}^A_B=
\begin{bmatrix}
    \textbf{R}^A_B & \textbf{o}^A_B\\
    0           & 1
\end{bmatrix},
\end{equation}
where $\textbf{R}^A_B$ is a $3\times3$ matrix representing the orientation (rotational) transformation from frame A to frame B, and $\textbf{o}^A_B$ is a $3\times1$ column vector representing the translation from the origin of frame A to frame B (both $\textbf{R}^A_B$ and $\textbf{o}^A_B$ are formulated with respect to frame A).  A series of such transformations can be performed to describe the total configuration of the robotic arm shown by

\begin{equation}
\textbf{H}^A_D=\textbf{H}^A_B\textbf{H}^B_C\textbf{H}^C_D,
\end{equation}
where the final result is a matrix describing the orientation and position of reference frame D with respect to frame A.

For Kresling modules at state (0), we employ the Jones kinematic model that describes the shape of these soft modules as a simple arc with constant curvature \cite{Jones2006b, Chawla2018a}. Thus, their corresponding transformation matrix is:

\begin{equation}
\textbf{H}_{(0)}=\
\begin{bmatrix}
    \cos\phi  & -\sin\phi \cos\theta & \sin\phi \sin \theta   & \kappa^{-1}\left(\sin \phi \left(1-\cos \theta\right)\right)\\
    \sin\phi  & \cos\phi\cos\theta   & -\cos \phi \sin \theta & -\kappa^{-1}\left(\cos \phi \left(1-\cos \theta\right) \right)\\
    0                & \sin \theta                   & \cos \theta                    & \kappa^{-1} \sin{\theta}\\
    0                & 0                                     & 0                                      & 1
\end{bmatrix},
\end{equation}
where 
\begin{equation}
\phi=\tan^{-1}\left( \frac{\sqrt{3}}{3}\frac{l_\text{ii}+l_\text{iii}-2l_\text{i}}{l_\text{ii}-l_\text{iii}}\right),
\end{equation}

\begin{equation}
\theta=2N\sin^{-1}\left[ \frac{\left( l_\text{i}^2+l_\text{ii}^2+l_\text{iii}^2-l_\text{i}l_\text{ii}-l_\text{ii}l_\text{iii}-l_\text{i}l_\text{iii} \right)^{1/2}}{3Nd}\right],
\end{equation}

\begin{equation}
\kappa=2 \frac{\left( l_\text{i}^2+l_\text{ii}^2+l_\text{iii}^2-l_\text{i}l_\text{ii}-l_\text{ii}l_\text{iii}-l_\text{i}l_\text{iii} \right)^{1/2}}{d\left(l_\text{i} +l_\text{ii}+l_\text{iii}\right)}.
\end{equation}
Here, $l_\text{i}$, $l_\text{ii}$, and $l_\text{iii}$ are the lengths of three driving tendons in this module, $N=1$, and $d$ $(= 32\text{ mm})$ is the distance between tendon and the longitudinal axis of Kresling module.  In the tests shown in Figure \ref{fig:RobotExp}(b,c), we use the motors to pull tendon i and release tendon ii and iii simultaneously so that $\Delta l_\text{{ii}} = \Delta l_\text{{iii}} = -0.5 \Delta l_\text{{i}}$.

For Kresling modules at state (1), we simply assume that they are straight links so that their corresponding transformation matrix is:

\begin{equation}
    \textbf{H}_{(1)}=
    \begin{bmatrix}
        1 & 0 & 0 & 0 \\
        0 & 1 & 0 & 0 \\
        0 & 0 & 1 & L_{(1)} \\
        0 & 0 & 0 & 1
    \end{bmatrix},
\end{equation}
where $L_{(1)}$ $(= 111.6\text{ mm})$ is the resting length of the Kresling module at the stable state (1).

\begin{figure}[]

    \centering \includegraphics[]{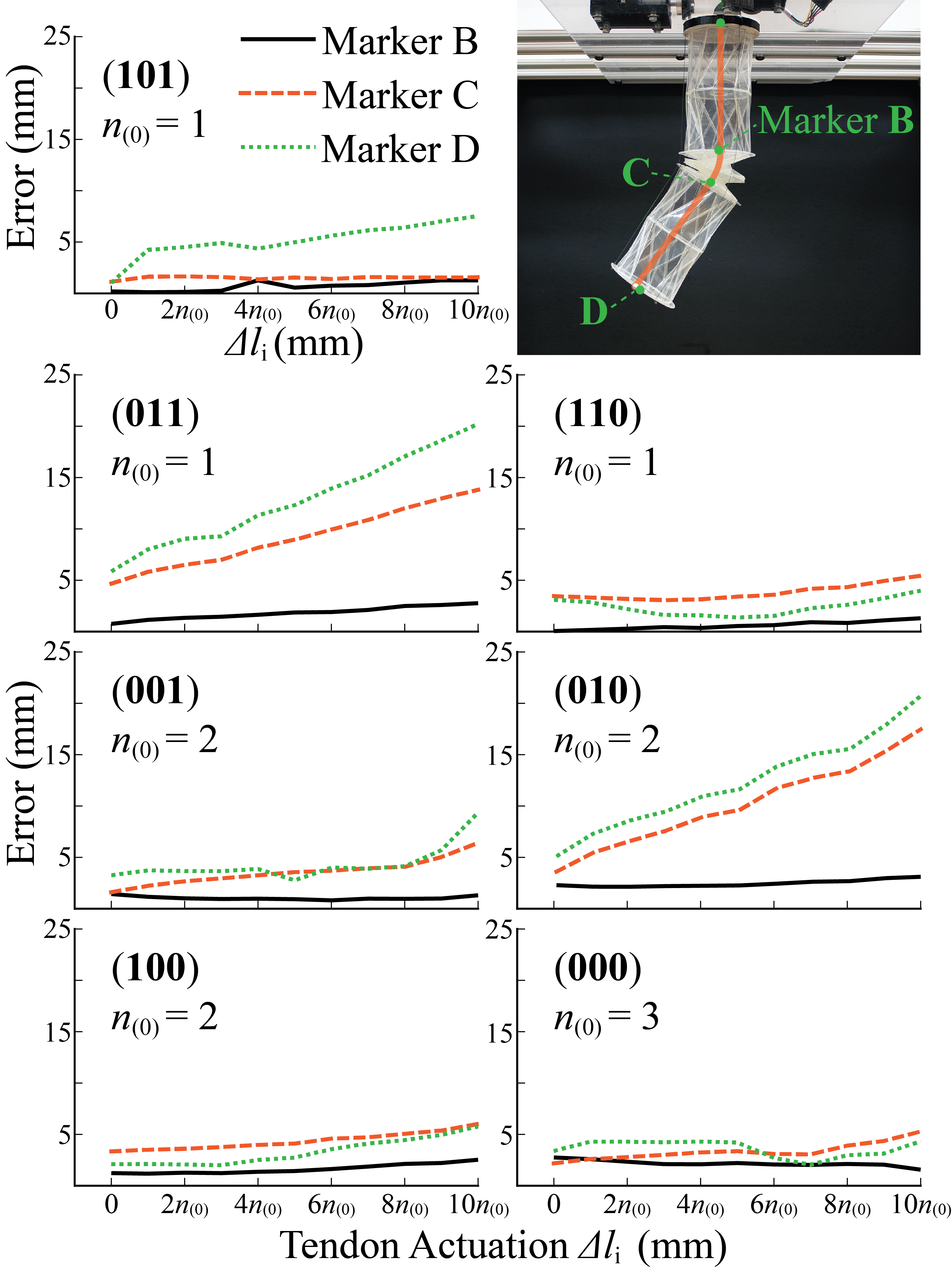}
    \caption{
    The differences between experimentally measured Markers B, C, and D positions and the corresponding kinematic model prediction. 
    }
    \label{fig:RobotErr}
\end{figure}

Overall, the robotic arm deformations agree well with the kinematic predictions at different tendon actuation levels.  Figure \ref{fig:RobotErr} further summarizes the differences between the experimentally measured marker positions and the corresponding predictions based on the kinematic model.  For most of the stable configurations, these differences are smaller than 5 mm, which is small compared to the overall length of the robotic arm (ranging from approximately 50 mm to 270 mm depending on the articulation setup).  These discrepancies probably originate from fabrication imperfections, gravity, as well as other small and complex Kresling deformations due to its compliant nature.  On the other hand, the position errors of Markers C and D are more significant at the (011) and (010) configurations as the robotic arm is displaced further.  In these two configurations, the Kresling module at the base behaves like a joint, which is immediately followed by a link-like module.  As a result, the error from the base joint amplifies and accumulates further down the chain.  Regardless, the linkage-like behavior is still evident, so these marker positions errors can be easily reduced by implementing feedback control.

\section{Conclusion}

This study proposes to exploit the bistability of Kresling origami to enable localized bending stiffness tuning and reconfigurable articulation of a modular robotic arm.  By strategically switching the Kresling modules between the stiff, link-like stable state (1) and soft, joint-like stable state (0), one can significantly reduce the effective degrees of freedom of a compliant manipulator to both simplify the control requirements and increase motion precision.  Via both analytical studies using a nonlinear bar-hinge method and experimental validations using a customized three-point bending method, we uncover the correlations between the magnitude of bending stiffness change and the underlying crease pattern design of the Kresling module.   Generally speaking, the Kresling modules with more polygon sides, lower resting length at contracted stable state, and higher angle ratio can offer a more significant and robust change in bending stiffness between their two stable states.  However, the module length at stable state (0) needs to be chosen carefully to ensure sufficient kinematic freedom for bending.  In experiments, the paper-based Kirigami modules achieve an order of magnitude change in bending stiffness by simply switching between two stable states, without the need for a continuous power supply to maintain this stiffness change.

We construct a tendon-driven, proof-of-concept robotic arm prototype by assembling three identical Kresling modules.   This robotic arm successfully validates the concept of reconfigurable articulation by exhibiting linkage-like deformations at different stable configurations.  Moreover, the magnitudes of these deformations agrees well with analytical predictions according to a kinematic model.   Therefore, the results of this study lay down the foundation for a reconfigurable robotic arm that can adapt to different manipulating task requirements with a significantly reduced control effort.

\section{Acknowledgement}
J. Kaufmann and S. Li acknowledge the support from Clemson University (via startup fund and Dean's Faculty Fellow Award) and the National Science Foundation (CMMI: 1751449 CAREER, and 1933124).

\bibliographystyle{elsarticle-num-names}
\bibliography{references.bib}

\newpage
\section*{Appendix 1: Fundamentals of the Bar-Hinge Model}

\begin{figure}[h]

    \centering \includegraphics[]{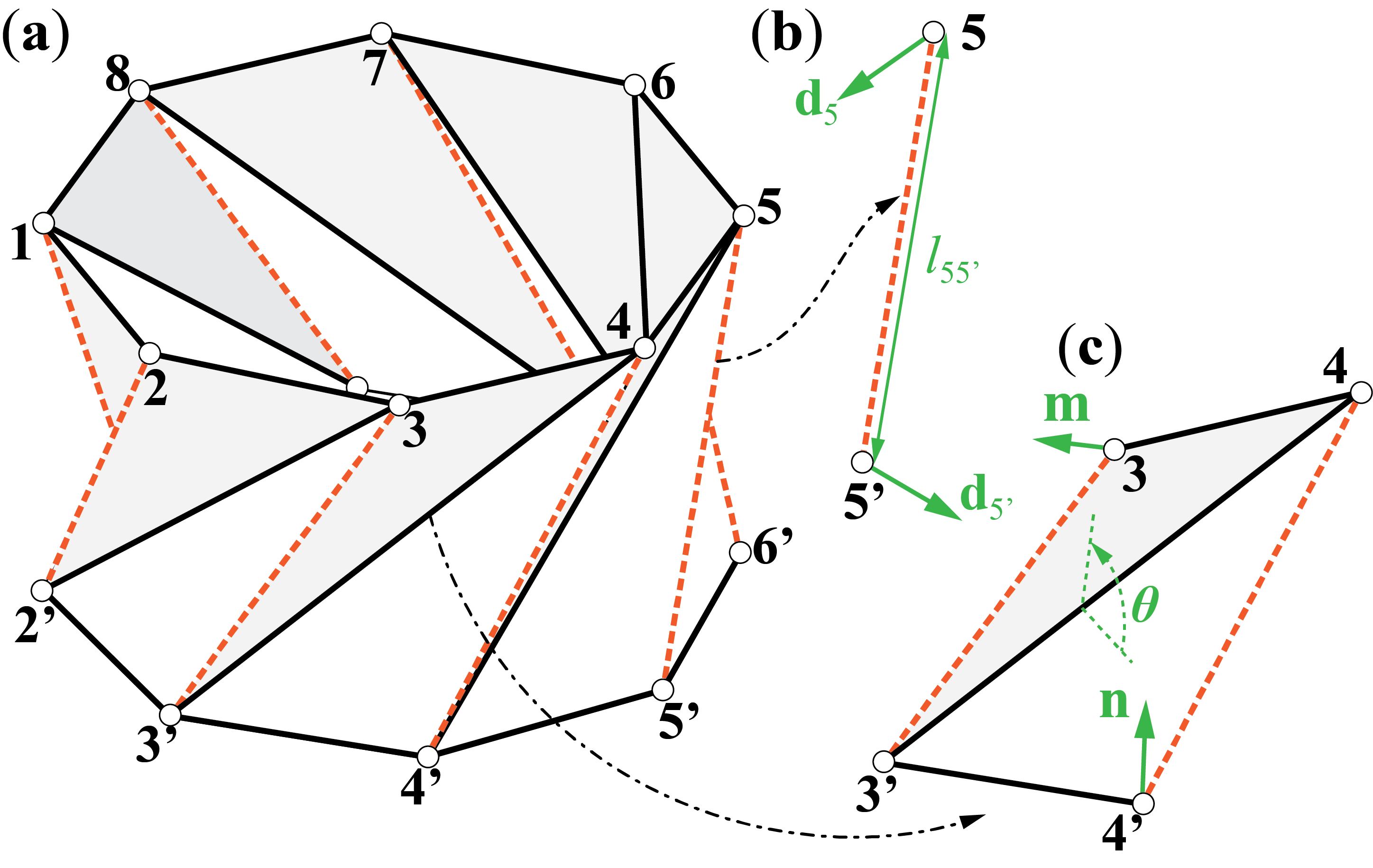}
    \caption{
    The kinematic setup of the bar-hinge method. (a) The overall bar-hinge setup of the Kresling origami cell. (b) Stretching of a bar element. (c) Folding of a crease along bar 3'-4.
    }
    \label{fig:BarHinge}
    
\end{figure}

The formulation of this appendix is adapted from a previous study by Liu and Paulino \cite{Liu2017c}, in which interested readers can find the details.   The overall stiffness of the bar-hinge system has two components. One comes from the stretching of the bar elements and the other from the folding (or bending) between adjacent triangular facets defined by these bar elements.  Using the bar element connecting pin-joints 5 and 5' as an example (Figure \ref{fig:BarHinge}(b)), one can define
$\mathbf{u}_{55'}=\left[\mathbf{d}_{5}^\intercal \;\; \mathbf{d}_{5'}^\intercal\right]^\intercal$, 
where $\mathbf{d}_{5}$ and $\mathbf{d}_{5'}$ are the displacement vector of the pin joint \#5 and \#5', respectively.  $l_{55'}$ is the length of this bar element.  The Green-Lagrangian strain of this bar element is

\begin{equation}
    \varepsilon_{55'}=\mathbf{B}\mathbf{u}_{55'}+\frac{1}{2}\mathbf{u}_{55'}^\intercal\mathbf{D}\mathbf{u}_{55'},
\end{equation}
where
\begin{equation}
    \mathbf{B} = \frac{1}{l_{55'}}\left[-\mathbf{e} \;\;\; \mathbf{e} \right], 
\end{equation}
\begin{equation}
    \mathbf{D} = \frac{1}{l_{55'}^2}
        \begin{bmatrix} 
            \mathbf{I}_{3\times3} &  -\mathbf{I}_{3\times3} \\ 
            -\mathbf{I}_{3\times3} &  \mathbf{I}_{3\times3}
        \end{bmatrix}.
\end{equation}

Here, $\mathbf{e}=[1 \;\; 0 \;\; 0]$, and $\mathbf{I}_{3\times3}$ is the identity matrix of size $3 \times 3$.  The tangent stiffness matrix components corresponding to this bar element are

\begin{equation}
\mathbf{K}_{55'}^\text{(bar)} =  k^\text{s}_{55'} l_{55'}\left(\textbf{B}^\intercal+\textbf{D}\mathbf{u}_{55'}\right)\left(\textbf{B}^\intercal+\textbf{D}\mathbf{u}_{55'}\right)^\intercal +f_{55'}l_{55'}\mathbf{D},
\end{equation}

\noindent where $k^\text{s}_{55'}$ is the axial rigidity of this bar element, and $f_{55'}$ is the resultant longitudinal force.  It is worth noting that this stiffness matrix involves both the linear term and nonlinear terms related to geometry and initial displacement \cite{Liu2017c}. One can then apply similar formulations to all bar elements and assemble the global bar stiffness matrix.

Besides bar stretching, folding and bending between the adjacent triangular facet are also crucial sources of stiffness.  Here, the creases behave like hinges with prescribed torsional spring stiffnesses.  Using the valley crease defined by pin-joints \#3' and \#4 as an example, one can calculate the dihedral angle between the two adjacent facets based on their surface normal vectors (e.g., $\mathbf{m}$ and $\mathbf{n}$ in Figure \ref{fig:BarHinge}(c)) so that

\begin{equation}
\theta=\eta \cos^{-1}\left( \frac{\mathbf{m}\cdot\mathbf{n}}{\|\mathbf{m}\|\|\mathbf{n}\|}\right),
\end{equation}
where the surface normal vectors $\mathbf{m}=\mathbf{r}_{33'}\times\mathbf{r}_{43'}$,  $\mathbf{n}=\mathbf{r}_{43'}\times\mathbf{r}_{44'}$, and $\eta$ is a sign indicator in that
\begin{equation}
\eta =
  \begin{cases}
    \text{sgn}(\mathbf{m}\cdot\mathbf{r}_{43'})       & \quad \text{if   } \mathbf{m}\cdot\mathbf{r}_{43'} \neq 0;\\
    1  & \quad \text{if   } \mathbf{m}\cdot\mathbf{r}_{43'} = 0.
  \end{cases}
\end{equation}

The elements of tangent stiffness matrix corresponding to this dihedral angle are are defined as 
\begin{equation}
K^{\text{(fold)}}_{3'4}=k^\text{f}_{3'4} l_{3'4} \frac{\text{d}\theta}{\text{d}\mathbf{x}}\otimes\frac{\text{d}\theta}{\text{d}\mathbf{x}}+m_{3'4}\frac{\text{d}^2\theta}{\text{d}\mathbf{x}^2},
\end{equation}
where $\otimes$ is tensor product, $l_{3'4}$ is the crease length corresponding to this dihedral angle, $k^\text{f}_{3'4}$ is the torsional spring stiffness \emph{per unit length} of this crease, $m_{3'4}$ is the resultant torque, and $\mathbf{x}$ is the position vector of the related pin-joints at the current configuration.

\section*{Appendix 2: Measuring the Crease Torsional Stiffness}

\begin{figure}[t]

    \centering \includegraphics[]{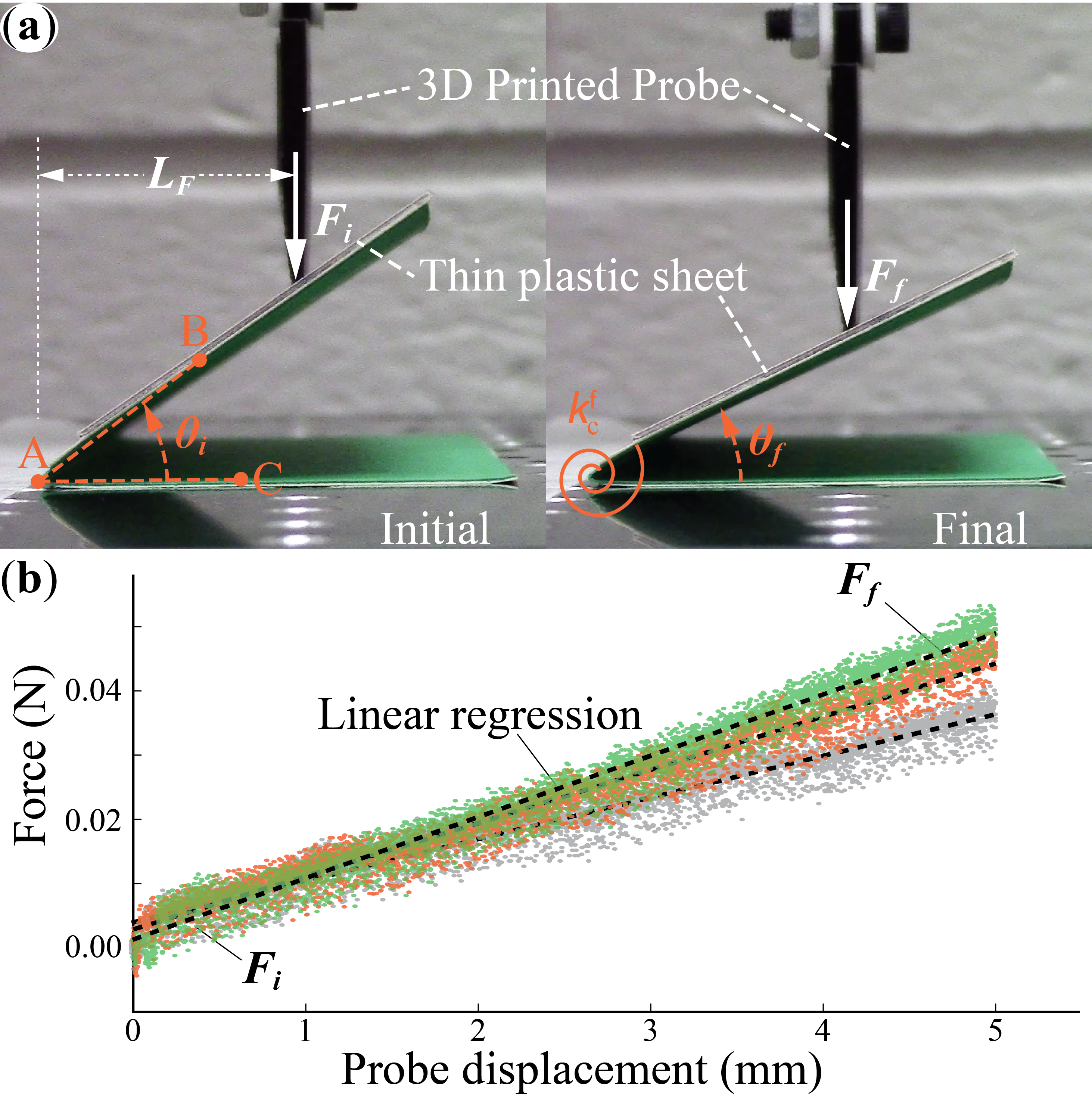}
    \caption
    {
    Experimental setup for measuring the crease torsional spring stiffness per unit length ($k^\text{f}_\text{c}$). (a) Screenshots for the video footage that shows the course of these tests. (b) Three sets of force-displacement data that shows consistent close-to-linear behaviors.
    }
    \label{fig:Crease} 
    
\end{figure}

We experimentally measure the crease torsional stiffness $k^\text{f}_\text{c}$ for the nonlinear bar-and-hinge model by using paper-based, hinge-like samples, each  consisting of two 15.24 cm by 4.45 cm rectangular facets connected by a perforated crease (Figure \ref{fig:Crease}).  The creases are fabricated on a Cricut Maker\textsuperscript{TM} cutting plotter.  We reinforce the upward-facing facet with a thin plastic sheet to eliminate panel bending and secure the bottom facet to the base plate of an ADMET eXpert\textsuperscript{TM} 5601 universal testing machine with double-sided tape. A 3D-printed, wedge-shaped probe is used to distribute compressive forces evenly across the upward facet (Figure \ref{fig:Crease}(a)). 

Before carrying out each test, we place the samples under a slight compressive load to ensure sufficient contact between the probe and sample. Then, a controlled probe displacement of 5 mm deforms the samples in the downward vertical direction (Figure \ref{fig:Crease}(a)).  We take high-resolution videos of the deformed samples during these tests and use MATLAB image processing to measure the sample deformation.  More specifically, we extract the first and last frames from the videos to represent the starting and ending configurations, respectively. Then we manually select the crease vertex point A, a point B on the upward facet, and a point C on the downward facet and then retrieve their respective x and y coordinates using an image processing program (Figure \ref{fig:Crease}(a)).  In this way, we can calculate the distance between these points and the corresponding dihedral folding angle $\theta$:
\begin{equation}
\theta=\cos^{-1}\left(\frac{L_{AB}^2+L_{AC}^2-L_{BC}^2}{2L_{AC}L_{AB}}\right)
\end{equation}
The measured force-displacement curve is close to linear (Figure \ref{fig:Crease}(b)). Therefore, we performed a linear regression to the data to estimate the initial force $F_i$ and final external force $F_f$, respectively.  In this way, the linear crease torsional spring stiffness \emph{per unit length} can be estimated as 
\begin{equation}
k^\text{f}_\text{c}=\frac{L_F\cos \theta_i \left(F_f-F_i\right)}{W\|\theta_i-\theta_f\|},
\end{equation}
where $\theta_i$ and $\theta_f$ are the initial and final dihedral angles, respectively. $L_F$ is the distance from the crease vertex to the applied force, and $W$ is the width of this crease sample. We fabricated five identical samples and conducted three load cycles on each sample.  The measured crease torsional stiffness per unit length is $k^\text{f}_\text{c}=0.047 \pm 0.011$ N/rad.  The averaged $k^\text{f}_\text{c}$ value is used in the nonlinear bar-hinge model.

\begin{equation}
K+\delta K^e=(k+\delta k^e) l \frac{\text{d}\theta}{\text{d}(\mathbf{x}+\delta\mathbf{x}^e)}\otimes\frac{\text{d}\theta}{\text{d}(\mathbf{x}+\delta\mathbf{x}^e)}+(m+\delta m^e)\frac{\text{d}^2\theta}{\text{d}(\mathbf{x}+\delta\mathbf{x}^e)^2},
\end{equation}
$$m=k(\theta-\theta_0)+m_l(t,p,v)$$
\end{document}